\documentclass[conference,letterpaper]{IEEEtran}

\usepackage[svgpath=./img/]{svg}

\usepackage{microtype}
\usepackage{graphicx}
\usepackage{subfigure}
\usepackage{booktabs} 
\usepackage{tabularx}
\usepackage{multirow}  
\usepackage{multicol}

\usepackage{mathtools}
\usepackage{amsmath} 

\usepackage{algorithm}
\usepackage{algorithmic}

\usepackage{hyperref}

\usepackage{cite}

\ifCLASSINFOpdf
\else
\fi

\begin{document}
%
\title{An efficient and flexible inference system for serving heterogeneous ensembles of deep neural networks}

\author{
\IEEEauthorblockN{Pierrick Pochelu}
\IEEEauthorblockA{
TotalEnergies SE\\
Pau, France\\
pierrick.pochelu@totalenergies.com}
\and
\IEEEauthorblockN{Serge G. Petiton}
\IEEEauthorblockA{Univ. Lille, CNRS, UMR 9189 CRIStAL \\
Lille, France\\
serge.petiton@univ-lille.fr}
\and
\IEEEauthorblockN{Bruno Conche}
\IEEEauthorblockA{
TotalEnergies SE \\
Pau, France\\
bruno.conche@totalenergies.com}
}


%

\IEEEoverridecommandlockouts
\IEEEpubid{\makebox[\columnwidth]{ 978-1-6654-3902-2/21/\$31.00 ©2021 IEEE \hfill} \hspace{\columnsep}\makebox[\columnwidth]{ }}

\maketitle

\begin{abstract}
Ensembles of Deep Neural Networks (DNNs) has achieved qualitative predictions but they are computing and memory intensive. Therefore, the demand is growing to make them answer a heavy workload of requests with available computational resources. Unlike recent initiatives on inference servers and inference frameworks, which focus on the prediction of single DNNs, we propose a new software layer to serve with flexibility and efficiency ensembles of DNNs.

Our inference system is designed with several technical innovations. First, we propose a novel procedure to found a good allocation matrix between devices (CPUs or GPUs) and DNN instances. It runs successively a worst-fit to allocate DNNs into the memory devices and a greedy algorithm to optimize allocation settings and speed up the ensemble. Second, we design the inference system based on multiple processes to run asynchronously: batching, prediction, and the combination rule with an efficient internal communication scheme to avoid overhead.

Experiments show the flexibility and efficiency under extreme scenarios: It successes to serve an ensemble of 12 heavy DNNs into 4 GPUs and at the opposite, one single DNN multi-threaded into 16 GPUs. It also outperforms the simple baseline consisting of optimizing the batch size of DNNs by a speedup up to 2.7X on the image classification task.

\end{abstract}

\begin{IEEEkeywords}
Neural network, ensemble learning, inference system
\end{IEEEkeywords}

Ensembles of deep neural networks are now well-known for producing qualitative predictions. First, ensembles of deep neural networks significantly improve generalization accuracy compared to one single model \cite{useoverfit:1995}. Second, they generally produce well-calibrated uncertainty estimates \cite{uncertainty:2017}.

Today, multiple researchers and practitioners have well understood the benefit of ensembling DNNs. For example, in cyber-attack detection \cite{enscyb:2020}, time series classification \cite{enstim:2018}, medical image analysis \cite{ensimg:2021}, semi-supervision \cite{enssem:2021} and unbalanced text classification \cite{enstex:2020}.  Further, several winners and top performers on challenges routinely use ensembles to improve accuracy.

It is also of common knowledge that a machine learning model creates value only in the inference phase. That is to say when the model is hosted on hardware and ready to receive input data samples from a client application and return their prediction. However, ensembles of DNNs are memory, time and computational resources expensive and no inference system is still adapted to optimize and to serve ensembles.

Two different software layers have been proposed to infer efficiently individual DNNs but does not propose to combine them: the inference servers (Triton \cite{tritonserv}, Ray Serve \footnote{\url{https://docs.ray.io/en/master/serve/index.html}} \cite{ray}
, Tensorflow Serving \cite{tfserv} and TorchServe \cite{torchserv}) serve the inference systems (such TensorRT \cite{tensorrt}, OpenVINO \cite{openvino}, ONNX \cite{onnx:} and TFLite \cite{tflite:}) predictions. Our work attempt to fill this gap between the current inference system technologies and the ensembles of deep neural networks.

The question we attempt to answer is simple but the solution is challenging \textbf{``How to systematically allocate an ensemble of DNNs to a given set of devices?''}. The systematic procedure must be endowed with two main qualities. Firstly, the flexibility, the systematic procedure aims to fit the ensemble in memory to be ready to answer requests, even if the number of devices is lower than the ensemble size. An ideal flexible solution must be able to allocate heterogeneous DNNs (such ResNet, Inception, EfficientNet, ...) on modern clusters containing heterogeneous devices (such CPUs, GPUs, TPUs, ...). Secondly, the efficiency, when an ensemble fits in memory it should optimize the usage of underlying multi-cores devices with minimum overhead due to data transfer.

We design the answer in three points.
\begin{itemize}
    \item First we propose the allocation matrix as the formalism of the decision space in an innovative way. It designs how DNNs are allocated into the devices including co-localization (multiple DNNs instances into the same device), data-parallelism (one DNN multi-threaded on multiple devices), and the batch size of each DNN instance (which controls devices cores usage, memory consumption, data exchange).
    \item We propose an \textit{allocation matrix optimizer}. It runs successively a worst-fit-decreasing algorithm to fit DNNs in memory and a bounded greedy algorithm to speed up the allocation. 
    \item Finally, we propose a design of the \textit{inference system}. To avoid overhead, it asynchronously runs: data batching, DNNs predictions, and the combination rule with an efficient internal communication scheme.
\end{itemize}

Our paper follows this structure:  (1) we show the recent progress in inference software. (2) We introduce our inference server for ensembles DNNs, how to use it and its internal mechanisms. (3) We perform multiple benchmarks and discuss the flexibility and the efficiency of our inference system.

\section{Inference software}

\subsection{Inference frameworks}

The inference frameworks contain generally two main functions, the ``load'' function to load a trained DNN from the disk to a targeted device and the prediction function $f(x) \rightarrow y$ with $x$ the data samples and $y$ the associated predictions. The most sophisticated inference frameworks \cite{tensorrt}, \cite{openvino}, \cite{onnx:}, \cite{tflite:} perform post-training optimization such operations optimization and device-specific optimization with low or no impact on the accuracy. We use here the Tensorflow deployment (``pb files'') as the underlying format and focus our work on the allocation challenge. 

A critical part of optimizing the performance of a DNN model is its batch size. It controls the internal cores utilization, the memory consumption, and the data exchange between the CPU containing input data and the device supporting the DNN (if different). That is why some tools \footnote{\url{"https://github.com/triton-inference-server/model_analyzer/blob/main/docs/config_search.md"}} scans multiple batch size values of a given DNN. Then, the batch size offering the best performance is used. This Best Batch Strategy (or BBS) is a relevant mechanism to optimize a single DNN on one device, but it is a naive and rigid method to optimize multiple DNNs predicting ensemble. Due to the lack of efficient technique to handle ensemble of DNNs in inference mode, we design an \textit{allocation matrix optimizer} that we compare to BBS.

\subsection{Inference servers}

The last few years, we show the emergence of software \cite{clipper:2017} \cite{tritonserv}, \cite{ray}, \cite{tfserv}, \cite{torchserv} to serve inference framework predictions as a service. Most of them wrap predictions in a REST service but other technologies exist such database management service \cite{dbms}. They implement often the same features. Ensemble selection allows the client application to choose the model which will answer among multiple applications or the same application but different trade-offs between accuracy and speed. To improve performance under redundant requests, caching allows avoiding recomputing similar requests. When the amount of requests is low and irregular, adaptative batching allows triggering prediction before the buffered batch is full to improve the latency.

Our work benefits from those inference server technologies. The novelty is that we handle heterogeneous ensemble of DNNs by adding an intermediate software layer between low level DNN inference frameworks and inference servers.

\section{The workflow}

\begin{figure}
        \centering
        \includegraphics[page=1,width=\linewidth,trim={0.8cm -2cm 9cm 0},clip]{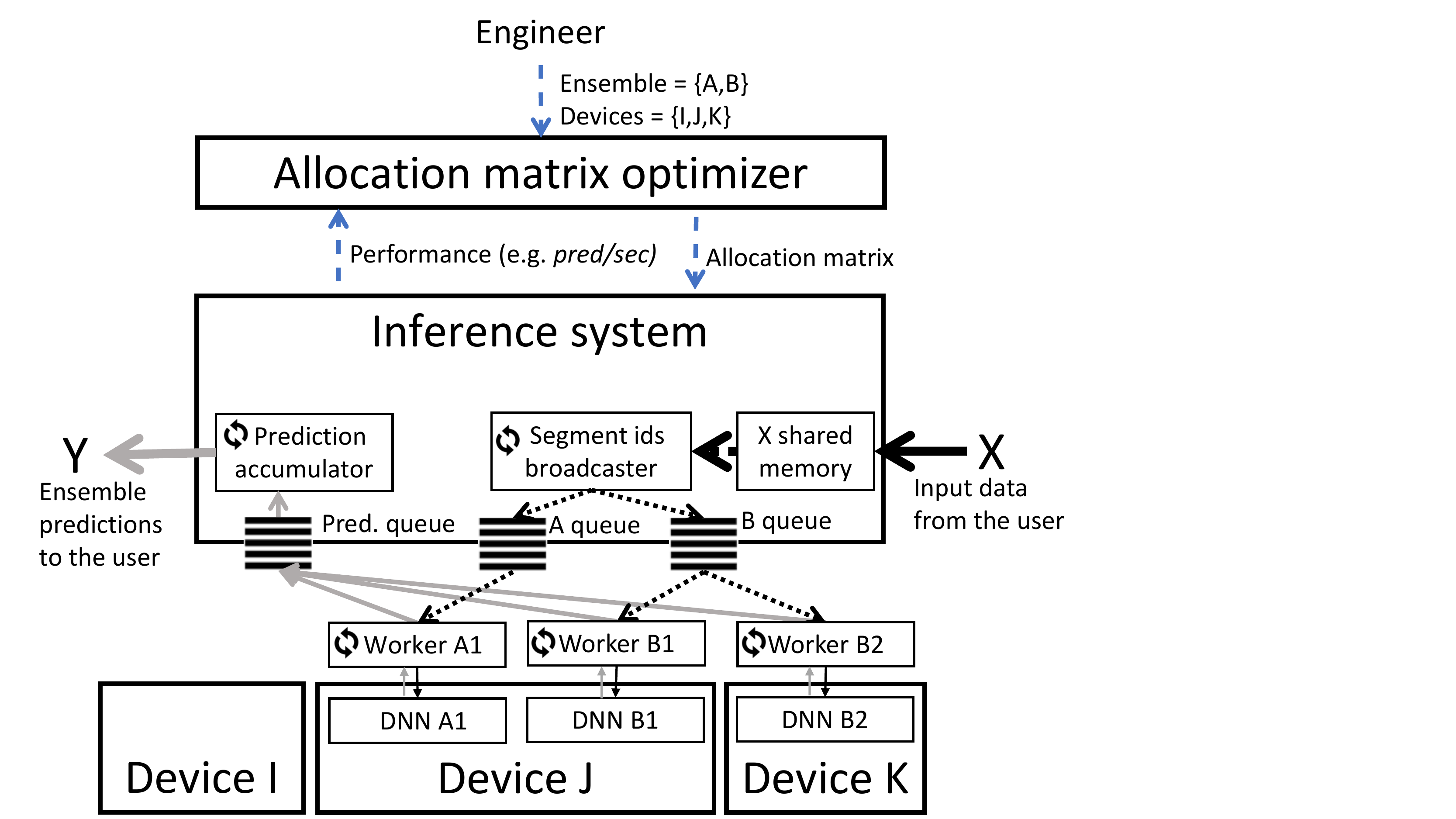}
        \includegraphics[page=3,width=\linewidth,trim={1cm 11cm 8cm 0},clip]{img/smart_server.pdf}
        
        \caption{Illustration of our inference server on a toy example of allocation of 2 DNNs into 3 devices. The DNN model B is run by 2 data-parallel workers on device J and device K. The DNN instances A1 and B1 are co-localized in device J. The corresponding allocation matrix is described in the bottom left corner. Threads inside a worker are not fully described for visibility purposes (see figure~\ref{fig:worker}).}
        \label{fig:alloc}
\end{figure}

Once an ensemble has been trained and built, we need to serve it efficiently on the available computing resources. We design an efficient inference pipeline illustrated in figure \ref{fig:alloc}. 

In the following sections, we will describe first how the inference server is used. Then, the allocation matrix which drives performance. Finally, main components of our inference server is presented one-by-one: the \textit{inference system core}, the \textit{worker-pool} and the \textit{allocation optimizer}.

\subsection{Using our inference server}

The engineer who is responsible for managing the inference server provides to the \textit{matrix allocation optimizer} the ensemble of DNNs, the CPUs, and the GPUs to use. Sometimes, the engineer does not want to give all available devices into its cluster to deploy an ensemble, so he can keep GPUs for other applications. The matrix allocation optimizer will automatically optimize the allocation of the ensemble for the given devices.


Once the allocation matrix is computed, the inference system is deployed online. It implements the usual inference server features such as an HTTP/HTTPS wrapper and adaptative batching. To be more precise, the term ``adaptative batching’’ can now lead to confusion. The buffer waiting request is now defined by the size of segments and not the batch size of the individual DNNs.

\subsection{The allocation matrix data structure}

The allocation matrix is a data structure describing with high flexibility the design of the \textit{worker-pool} used by the \textit{inference system}.  It designs how DNNs are allocated into the devices including co-localization (multiple DNNs instances into the same device), data-parallelism (one DNN multi-threaded on multiple devices), and the batch size.

In this matrix, element 0 means an absence of a worker process in the given device, while other values represent the batch size. The workers co-localized into the same device are readable as non-zero values in the row of the allocation matrix. The workers which are instances of the same DNN are data-parallel, they are readable as the non-zero values in a given column.  

Finally, some devices may not be used, so we can observe rows containing only zero values. But it is illicit to have a column with only zero values. In other words, all DNNs must be represented in the ensemble.

\subsubsection{Co-localization}

Co-localization allows fitting more DNNs than the number of devices (the pigeon-hole principle). It may allow also maximize the utilization of internal cores into a device too.

It is well known that the batch size is an important setting. However, optimal batch size values in the context of multiple DNNs co-localized into one GPU are challenging to find. In general, the larger batch may increase cores utilization and it consumes more memory. Due to those complex relationships between multiple DNNs co-localized and hardware, only benchmarks allow knowing the performance of co-localized models. And more, only multiple benchmarks allow finding good settings such as which DNNs should be put together and their optimal batch size values.

\subsubsection{Data-parallelism}

Data-parallelism allows to speed up a prediction of one DNN using multiple devices. The workers run the same DNN but in different instances. They take data samples to predict from the same input FIFO queue.

Ideally, $n$ threads should multiply the number of images predicted per second by $n$. Yet, because we use multiple shared data structures: one segment ids FIFO queue, the shared data memory, and the prediction FIFO queue, perfect scalability is not ensured. Only benchmarks allow to compute the performance and if increasing the number of workers is worth it.

\subsection{The inference system}


The inference system is the core component of the server. It is a function $f(X, A)\mapsto\{Y, S\}$ with $X$ data samples to predict and $Y$ the associated predictions of the ensemble of DNNs. $A$ is the allocation matrix driving the worker-pool construction and $S$ is the performance score. 

The inference system can be run with 2 modes. In ``Deploy Mode'' it is deployed online to serve client requests, $A$ is fixed and $S$ is ignored. In ``Benchmark Mode'' it measures the performance $S$ provided by the allocation matrix $A$ on the data calibration samples $X$, and $Y$ is ignored.

To accelerate the inference system we design it with multiple processes. The \textit{segment identifiers broadcaster} which splits the incoming workload of requests into segments, the \textit{worker pool} containing DNNs instance and return segments of predictions in the \textit{combination accumulator} FIFO queue, the \textit{combination accumulator} combines segments of predictions and returns to the client the final prediction: the prediction of the ensemble.


\subsubsection{The segment ids broadcaster} 

The \textit{inference system} contains thread-safe FIFO queues allowing to broadcast and gather information with the workers. It contains also the \textit{X shared memory}, this is a heavy buffer of data readable by all the workers. All those data structures are stored in the RAM.

To avoid transmitting heavy messages and stressing thread-safe FIFO queues, all images information passing through queues are gathered into segments. All segments contain  $N$ samples, except the last segment which contains the information of the remaining samples. 

After getting a segment identifier $s$, such as $s\geq0$, a worker knows he is responsible to predict the images from $start(s)=s*N$ position to the $end(s)=min((s+1)*N, nb\_images)$ position with $nb\_images$ the number of images in the \textit{X shared memory}. The \textit{segment ids broadcaster} can also put special values like $s=-1$ to ask workers to shut down before terminating the overall inference system.

For example, in the figure \ref{fig:alloc} if the user requests the prediction for 300 images with $N=128$, they are represented internally as 3 segments, two are size 128 and one is size 44. Then, the \textit{segment ids broadcaster} put 6 messages: 0, 1, 2 integers into A queue and B queue.

\subsubsection{The prediction accumulator} 

This process combines efficiently predictions from workers and when it is finished it returns them to the client.  After predicting a segment of data, a worker puts a message in the prediction queue. Each of these messages is a triplet $\{s, m, P\}$ with $s$ the segment identifier, $m$ the model identifier and P the prediction matrix of dimension $(end(s)-start(s)) \times C$ and $C$ the prediction length for each image e.g., the number of classes. To combine predictions, the \textit{prediction accumulator} first allocates a buffer $Y$ of dimension $nb\_data \times C$ zeroed. Then, it updates the cumulative prediction each time it receives a message triplet $\{s, m, P\}$. The Python code using Numpy arrays of the averaging accumulation is simply:
\begin{verbatim}
Y[start(s):end(s)]+=P/M
\end{verbatim}

With $M$ the number of DNNs in the ensemble. Other combination rules can be easily implemented such as majority voting or weighted averaging. Furthermore, other applications can require specific combination rules such as those applied in object detection \cite{WBF}. Any combination rule code must be developed by keeping in mind that predictions come into messages to be asynchronous with the neural network predictions.

To go into further details, the workers can send special messages to the \textit{prediction accumulator}. The special message \{-1, None, None\} allows notifying that a device has not enough memory to load or initialize a DNN. This triggers the shutdown of the \textit{inference system} and every process into it. The special message \{-2, None, None\} allows notifying that one worker is ready to serve after its initialization. Starting the inference system takes a few seconds, we know the \textit{inference system} is fully initialized and ready to receive the user requests when all workers send \{-2, None, None\} to the \textit{prediction accumulator}.


\subsection{The worker pool}

The worker pool gets segments from FIFO queues, predicts them, and returns the predictions to the \textit{prediction accumlator}.

\begin{figure}
        \centering
        \includegraphics[page=2,width=1\linewidth,trim={0cm 6.5cm 17cm 0.5cm},clip]{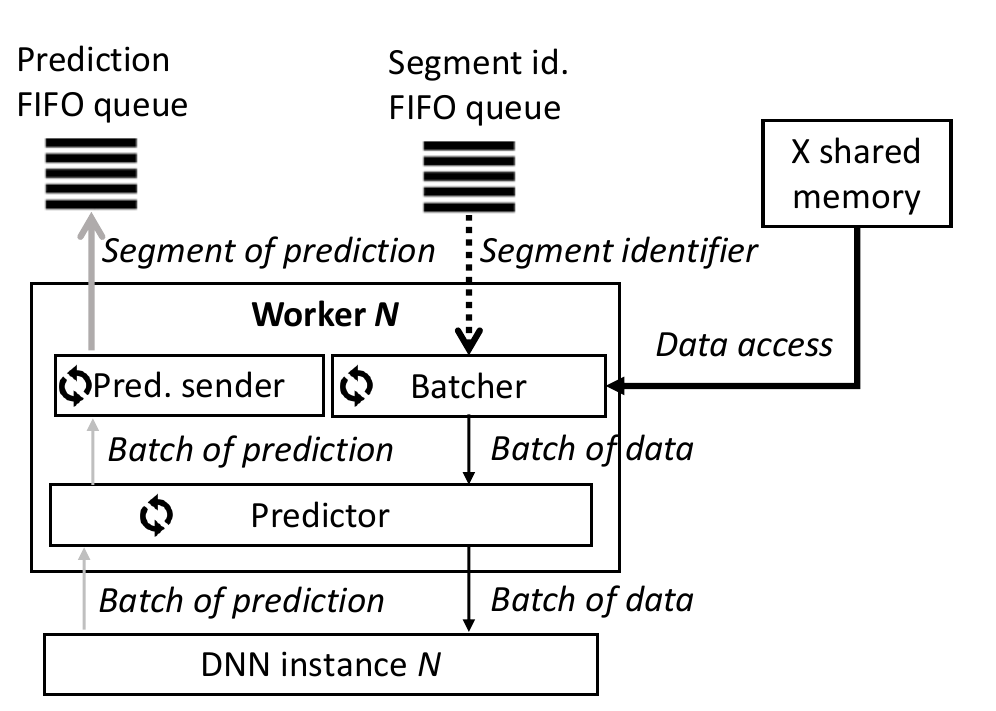}
        \caption{Anatomy of a worker. The meaning of arrows is the same than figure~\ref{fig:alloc}}
        \label{fig:worker}
\end{figure}

All workers are designed the same, they are illustrated in figure~\ref{fig:worker}. To be performant it contains 3 asynchronous threads: the \textit{batcher}, the \textit{predictor} and the \textit{prediction sender}.
\begin{itemize}
    \item The \textit{batcher} waits for incoming segment identifiers from the input FIFO queue. When it receives one, it splits the segment into a batch of data and gives them to the \textit{predictor}. Each worker has its batch size described in the allocation matrix. 
    \item The \textit{predictor} persists the DNN into the device memory. When it receives a segment identifier, first it loads the segment of data. Then it runs the prediction on each batch and finally, it gives the batch of predictions to \textit{the prediction sender}. 
    \item The \textit{prediction sender} gathers predictions batch by batch to build segments of prediction. When a segment is completed, this process puts the segment of prediction into the prediction FIFO queue.
\end{itemize}

\subsection{The allocation optimizer}
\label{sec:alloc}

The \textit{allocation matrix optimizer} goal is first to fit DNNs into the memory and then optimize their performance. That is why it runs first a worst-fit-decreasing algorithm to solve the bin-packing problem to fit DNNs into the memory of the device. Second, a greedy algorithm assesses thousands of matrices to find the faster one based on offline benchmarks. Finally, the best matrix is cached to avoid recomputing it again when the server will be restarted. These two algorithms are next detailed.





\subsubsection{Algorithm 1 - Worst-Fit-Decreasing with priority to GPUs}


It solves a bin packing problem to put objects (DNNs) into a finite number of bins (devices). All DNNs are set with the minimum batch size value (8 in our experiments). It is more exactly a bin packing with offline heuristics problem because the DNNs are known before execution and can be sorted in decreasing order. We attempt to find a feasible solution with the already known Worst-Fit Decreasing algorithm. 

This algorithm has the already known properties to optimize memory allocation. At each step of the worst-fit algorithm, it chooses to allocate a DNN into the device with the largest available memory. Therefore it may remain large memory space into the GPU after a DNN allocation. This remaining memory space can be big enough so that other smaller DNNs can also be placed in that remaining memory and therefore maximize the memory filling. Second, ordering the input list by decreasing DNNs size has been proved \cite{worstfit} to maximize the memory filling compared to a non-ordered version.
Finally, Worst-Fit prioritizes an equitable workload between homogeneous devices based on the memory criterion. On the opposite, First-Fit, Best-Fit and Next-Fit, attempt to fill the first devices and keep the last devices empty.

To improve significantly the speed up of the allocation, we hard code a rule to allocate in priority the GPUs rather than CPUs. It is of common knowledge that GPUs can run DNNs an order of magnitude faster than CPUs. Indeed, the CPUs start to be used by algorithm~\ref{algo:alloc} only when no more space is available on the GPUs.

\begin{algorithm}[h]
\caption{Worst-fit-decreasing with a priority to GPUs}
\label{algo:alloc}
\small
\begin{algorithmic}[1]

\STATE{\textbf{input:} $M$ the list of DNN models in the ensemble, $D$ the device set,  $default\_batch\_size$}
\STATE{\textbf{output:} $A$ the allocation matrix containing all models placed}
\STATE{\textbf{start}}

\STATE{ $A_{m,d}$ $\gets$ 0, $m=1 ... card(M)$ and $d=1 ... card(D)$}

\STATE{$M$ sorted in desc. order of memory size}

\FOR{$m$ in $M$}

\STATE{}
\STATE{// AG is the matrix where $m$ is put on the GPU-side}
\STATE{$g\gets$ more\_remaining\_memory($A$,$default\_batch\_size$,`GPU')}
\STATE{$AG$ $\gets$ copy($A$)}
\STATE{$AG_{m,g}$ $\gets$ $default\_batch\_size$}

\STATE{ }
\STATE{// AC is the matrix where $m$ is put on the CPU-side}
\STATE{$c\gets$ more\_remaining\_memory($A$,$default\_batch\_size$,`CPU')}
\STATE{$AC$ $\gets$ copy($A$)}
\STATE{$AC_{m,c}$ $\gets$ $default\_batch\_size$}

\STATE{ }
\STATE{// Put the model m}
\IF{fit\_mem($AG$)}
	\STATE{$A \gets AG$ // Priority to the GPU-side}
\ELSIF{fit\_mem($AC$)}
	\STATE{$A \gets AC$}
\ELSE
	\STATE{Error no device have enough memory}
\ENDIF

\ENDFOR
\STATE{return $A$}

\end{algorithmic}
\end{algorithm}

\texttt{more\_remaining\_memory} function returns the device with the most remaining memory. \texttt{fit\_mem} returns if the allocation matrix is feasible in terms of memory availability.

\subsubsection{Algorithm 2 - Bounded greedy optimization}

The algorithm 2 goal is to refine the starting allocation matrix and return a faster allocation matrix. Before explaining this algorithm, we will first provide a detailed analysis of the complexity of its decision space which drives our choices to design it.

The number of matrices is given by equation~\ref{eq:combi} with $D$ devices, $B$ batch size values and $M$ DNN models in the ensemble.
\begin{equation}
    total\_matrices=((B+1)^{D}-1)^{M}
    \label{eq:combi}
\end{equation}
The number of possible element values is $(B+1)$ all possible batch size values plus the 0 elements for no worker. The number of possible values in a single column is described by $(B+1)^{D}-1$. The term $(B+1)^{D}$ allows to brute-force all possible values in a column minus the zero columns which are forbidden. 

Note the two power terms which show the explosion number of combinations. For example, if we have 8 DNNs, 4 GPUs, and 1 CPU $total\_matrices\approx1.3E31$, this is much more than the number of stars in the observable universe. To put the problem into perspective, one allocation matrix takes an average of 40 seconds to be assessed. It includes the construction of the \textit{inference system} time plus the offline benchmark time. That is why only a few hundred matrices can be reasonably assessed by an optimizer.

A greedy algorithm provides an often effective solution to many combinatorial optimization problems. When applied to our case, it starts with the matrix given by algorithm~\ref{algo:alloc} and at each iteration, it assesses all neighborhood matrices and replaces the current matrix with the best-assessed matrix. We consider that two matrices are neighborhoods if they are both valid (again, no 0 columns) and if there is only one different element between them. The greedy algorithm is stopped when no neighborhood improves the current matrix.

The greedy algorithm breaks down the overall complexity described in equation~\ref{eq:combi} into a succession of combinations (or neighbors) measured by equation~\ref{eq:neigh} to assess. With $F$ the number of forbidden matrices we cannot explore such $0 \leq F \leq D$.
\begin{equation}
    total\_neighs=(B+1)*(D*M)-F
    \label{eq:neigh}
\end{equation}

Let's take our previous example, 8 DNNs, 4 GPUs, and 1 CPU. The total number of combinations is $total\_matrices\approx1.3E31$ but the greedy algorithm needs to assess only between 232 and 240 neighbors at each iteration.

Greedy algorithms are well-known approximation algorithms. However, the number of neighbors can still be very computing-intensive to evaluate in some cases, and more the number of required iterations can be large before finding an optimum. To limit the computing cost of the greedy we propose two bounds. First, each iteration evaluates at most $max\_neighs$ randomly drawn neighbors (line 9). We also limit the number of iterations to $max\_iter$ (line 6).

Despite that our algorithm is an approximation of a greedy algorithm, we have the guarantee that in the worst-case scenario a solution as good as the starting one is returned. This characteristic is inherited from the greedy algorithm. Line 18 can be read ``if we do not improve strictly the performance, the algorithm is stopped''.

The rate of visited neighbours is measured with $  \frac{max\_neighs}{total\_neighs}$. If it is close to zero it means the final performance may be very volatile. At the opposite,  $\frac{max\_neighs}{total\_neighs} \geq 1$ means all neighbours are visited and does not introduce volatility. Another source of volatility is the benchmark function (lines 4 and 11) but in practice, when the amount of calibration data samples is large enough, the measurement is stable.

The pseudo-code is presented in bloc code~\ref{algo:speed}.  \texttt{bench} function instantiates the pipeline with the given allocation settings (first argument) on the calibration data samples (second argument) and returns the performance to maximize or 0 if a DNN instance does not fit in memory.

\begin{algorithm}[]

\caption{Bounded greedy algorithm}
\label{algo:speed}
\begin{algorithmic}[1]

\STATE{\textbf{input:} $max\_iter$ maximum number of greedy iteration, $max\_neighs$ maximum number of neighboors to evaluate at each iteration, $A$ is the zeroed allocation matrix, $calib\_data$ contains calibration data to perform offline benchmarks}
\STATE{\textbf{output:} The optimized allocation matrix $A$}
\STATE{\textbf{start}}

\STATE{ $A\_speed$ $\gets$ bench($A$,$calib\_data$)}
\STATE{$iter$ $\gets$ 0}

\WHILE{$iter < max\_iter$}

\STATE{$neighs$  $\gets$ neighborhood($A$) }
\IF{length($neighs$) $>$ $max\_neighs$}
\STATE{$neighs$  $\gets$ draw randomly $max\_neighs$ samples from $neighs\_A$ }
\ENDIF

\STATE{$best\_A$, $best\_speed$  $\gets$ for all $n$ in $neighs$ return the best one and its score based on the bench($n$,$calib\_data$) criterion }
\IF{$best\_speed > A\_speed$}
    \STATE{$A$ $\gets$ $best\_A$}
    \STATE{$A\_speed$ $\gets$ $best\_speed$}
    \STATE{$iter$ $\gets$ $iter$ + 1}
\ELSE
    \STATE{// local maxima (or plateau) detected}
    \STATE{$iter$ $\gets$ $max\_iter\_greedy$}
\ENDIF

\ENDWHILE

\STATE{return  $A$}

\end{algorithmic}

\end{algorithm}

\section{Offline benchmark settings}
\label{sec:appdesign}

This section provides a detailed description of our experimental settings. Reading this section is not needed to understand the next sections.

\textbf{The performance metric.} The transmission of online messages with the client and their characteristics is application dependant and thus we rather perform offline benchmarks to evaluate the core prediction performance. This allows us to stay focused on the specificity to deploy ensembles under a heavy workload of requests. Therefore, our metric will be the throughput measured by the number of data samples predicted per second.

\textbf{Code and framework.}  All the asynchronous objects are implemented with the ``Multiprocessing'' built-in Python 3.6 package. Our FIFO queues, shared memory, and processes are respectively implemented with Queue class, Manager class, and Process class.

Additionally, we use two well-known external packages: the Numpy numerical library and Tensorflow 1.14 to deploy models widely supported by GPUs and CPUs.



\textbf{The hardware.} All results reported are performed into an HGX cluster containing 16 Tesla-V100 GPUs. The flexibility and efficiency of our server are analyzed by varying the number of GPUs to use from 1 to 16.

We also performed a few benchmarks on a computing node identical to the Oak Ridge Summit node. Both clusters have the same GPUs but different CPUs and different operating systems. The measured results are very similar to HGX so we decided do not to report the performances.

\textbf{The ensembles of DNNs.} To assess the efficiency and the flexibility of the inference system, we benchmark 5 heterogeneous ensembles named according to the number of models and the name of the database they learned from.

First, we build arbitrary 3 ensembles of present or past state-of-art DNNs. They have also been chosen for reproducibility purposes and they are easily implementable for any deep learning framework. IMN1 contains only ResNet152 and it allows us to show our workflow is general enough to optimize one single DNN. IMN4 contains 4 DNNs \{ResNet50, ResNet101, DenseNet121 and VGG19\}. IMN12 contains all DNNs from IMN1 and IMN2 plus \{ResNet18, ResNet34, ResneXt50, InceptionV3, Xception, VGG16, MobilNetV2\}. 

Then, we generate 2 other ensembles named FOS14 and CIF36. FOS14 contains 14 DNNs and is used for our in-house applications with 224x224 RGB images as input and predict among 91 classes. CIF36 is an ensemble of 36 DNNs to classify the images from the CIFAR100 dataset taking 32x32 RGB images and 100 classes.

FOS14 and CIF36 have been built with an in-house AutoML \cite{pochelu:2021} with a posthoc ensembling method to automatically build ensembles to maximize the prediction quality.  FOS14 and CIF36 are built around the Resnet skeleton from 10 to 132 layers and the number of filters in each convolution is multiplied from 0.5 to 3 compared to the usual ResNet architectures. We do not provide further details on their construction and training to stay focused on the inference phase.

\textbf{Calibration data samples.} The meaning of the data has no impact on any performance measured on the classification task. However, applying this server to other DNN methods may need a few realistic data samples to measure a relevant performance time. For example, in the Faster-RCNN \cite{FastRCNN:2015} architectures the RPN module iterates on some region of interest for each input image.

\textbf{The possible batch size values}. We fixed \{8, 16, 32, 64, 128\} as possible batch size values. More values increase significantly the number of possible allocation matrices and make the exploration of combinations more difficult. Fewer values reduce the degree of freedom to control internal parallelism, memory consumption, and data exchange.

\textbf{The segment size}. We evaluate multiple segment sizes and we observe that smaller values reduce the granularity of the workload and improve its distribution between processes. In all this work the segment size is fixed to 128. It should generally be equal to or greater than the maximum batch size.

\textbf{Algorithm 2 - Greedy allocation}. We choose $max\_neighs=100$ and $max\_iter=10$. This is a total of at most 1000 combinations to assess. On average, one matrix evaluation takes 40 seconds therefore those two chosen values limit the computing cost to 12 hours.

When $D-M>max\_iter$ with $M$ DNNs and $D$ devices. The $max\_iter$ is replaced with $D-M$. It allows getting a chance of using all devices when the number of devices is large. In all our benchmarks, it is used only three times with IMN1 ensemble on 12 GPUs and 16 GPUs and IMN4 on 16 GPUs.

\section{Offline benchmarks}

This section describes 3 performance analyses of our system. The first one estimates the overhead introduced by the inference system. Then we analyze the performance varying the ensembles and the number of GPUs. Finally, we will compare our allocation matrix optimizer to a simple baseline named ``Best Batch Size’’.

\subsection{Overhead of the inference system}

In all our benchmarks, the inference system overhead is measured at most 2\% of the total inference time. To estimate this overhead, we temporarily replace all the DNNs calls with a fake prediction containing only zero values, thus the \textit{prediction accumulator} still gathers predictions but returns zero values. 

We perform multiple offline benchmarks of this fake inference system and we report it takes at most 0.035 seconds in the case of IMN12 ensemble on 16 GPUs, in this case, \textit{allocation matrix optimizer} producing 22 workers. In comparison, the true inference system (without faking predictions), takes 2.528 seconds to predict 1024 images (i.e., throughput=405 in the table ~\ref{tab:servcifar}).

\subsection{Varying GPUs and ensembles}

Table \ref{tab:servcifar} shows the performance in terms of prediction performance of the 5 ensembles. The first major observation is that to serve efficiently an ensemble of DNNs we do not need systematically as much as GPUs as DNNs, but more GPUs generally improve performance.

\begin{table}[h]
\small
\addtolength{\tabcolsep}{-1.5pt}
\begin{tabularx}{\linewidth}{l|ll|ll|ll|ll|ll}
\toprule
~ & \multicolumn{2}{c}{IMN1} & \multicolumn{2}{c}{IMN4} & \multicolumn{2}{c}{IMN12} & 
 \multicolumn{2}{c}{FOS14} & \multicolumn{2}{c}{CIF36}  \\

\#G & A1 & A2  & A1 & A2 & A1 & A2 
& A1 & A2 & A1 & A2 \\
\midrule
1  & 106 & 136     & -  & -  & - &  -
& - & - & - & - \\
2  & 106 & 270     & 13 & 101 & - & - 
& 213 & 233 & - & - \\
3  & 106 & 394     & 158 & 199 & - & -  
 & 308 & 339 & - & - \\
4  & 106 & 539     & 160 & 251 & 15 & 24 
 & 380 & 410 & - & - \\
5  & 106 & 617  & 160 & 294 & 65 & 106 
 & 388 & 461 & 15 & 15 \\
6  & 106 & 722     & 160 & 351 & 103 & 194 
 & 397 & 470 & 35 & 37 \\
8  &  106 & 974 & 160 & 472 & 103 & 226  
 & 483 & 518 & 239 & 243 \\
12  & 106  & 1436 & 160 & 686 & 103 & 317 
 & 511 & 545 & 428 & 481 \\
16  & 106  & 1897 & 160 & 877 & 103 & 405
& 511 & 559 & 563 & 633  \\
\bottomrule
\end{tabularx}
\caption{We benchmark the throughput of 5 ensembles on different numbers of GPUs (+1 CPU). The mention `-' means out of memory error is returned. Because A2 is a stochastic algorithm, each run time was performed 3 times and the median value is reported.}
\label{tab:servcifar}
\end{table}

The \textit{allocation optimizer} successes in automatically constructing allocation matrices in the majority of assessed scenarios. It found a feasible solution to store 12 DNNs into only 4 GPUs. On the opposite, it also shows success to leverage multiple GPUs, for example, the Resnet152 model alone gets a Weak Scaling Efficiency of 87\% with 16 GPUs. Finally, when we compare algorithm 1 (Worst-Fit Decreasing) alone and algorithm 1 followed by algorithm 2 (Bounded Greedy) throughput, we observe also that the second algorithm produces generally a significant speed-up confirming the usefulness of this second algorithm.

We observe from the different matrices produced. To illustrate the decision making of the \textit{matrix allocation optimizer}, we show an example of a matrix allocation in figure~\ref{tab:imn4}.

\begin{table}
\begin{tabularx}{\linewidth}{lcccc}
\toprule
~ & ResNet50 & ResNet101 & DenseNet121 & VGG19 \\
\midrule
CPU  & 0        & 0         & 0           & 0     \\
GPU1 & 8        & 8         & 0           & 0     \\
GPU2 & 0        & 128       & 0           & 0     \\
GPU3 & 0        & 0         & 8           & 0     \\
GPU4 & 0        & 0         & 0           & 8    \\
\bottomrule
\end{tabularx}
\caption{Allocation matrix of IMN4 on 4 GPUs returned by our allocation matrix optimizer}
\label{tab:imn4}
\end{table}

In general, slower DNNs responsible for the performance bottleneck of the ensemble are multi-threaded in priority and their batch size optimized. Furthermore, we observe that when some models are co-localized their batch size is often chosen smaller. Finally, when we increase the number of GPUs, algorithm 2 automatically stops using the CPU. Indeed, the CPU is only used by Algorithm 1 and Algorithm 2 when the GPUs memories are full.

In addition to measuring the throughput, we also measure the good stability of this throughput for any allocation matrix. This is a desirable property for some industrial applications where the server must guarantee a certain quality of service. More formally, the function bench($A$,$fake\_data$) (algorithm 2) returned values varies to a relative standard deviation (RSD) below 2\% for any $A$. We also measure that when the rate $max\_neighs/total\_neighs$ is low (e.g., inferior to 0.2) the bounded greedy algorithm can return diverse matrices performance at different run-time until RSD=16\%.

\subsection{Baseline comparison}

We compare in table~\ref{tab:matrix} our \textit{allocation matrix optimizer} with the commonly used strategy we named BBS as baseline. The BBS uses $n$ GPUs for $n$ models and for each model it searches for the optimum batch size. It requires the same amount of GPUs as DNNs, this is a major limitation that requires small ensembles or it requires large hardware investment.

\begin{table}[h]
\small
\addtolength{\tabcolsep}{-2pt}
\begin{tabularx}{\linewidth}{lllllc}
\toprule
~ & \multicolumn{2}{c}{BBS baseline}  &  \multicolumn{3}{c}{Our server} \\
~ & img/sec & \#bench & img/sec & \#bench \\
\midrule
IMN1 / 1GPU  & 136 & 5 & 136 & 69  \\
IMN4 / 4GPUs  & 211 & 20 & 251 & 200  \\
IMN12 / 12GPUs  & 136 & 60 & 338 & 1000  \\
''  & '' & '' & 376 & 2000  \\
\bottomrule
\end{tabularx}
\caption{Comparison between two allocation strategies: The simple preferred batch size strategy and our proposed allocation matrix optimizer with different ensembles (from IMN1 to IMN12) and different GPUs (+1 CPU each time). Those two strategies produce an allocation matrix for the inference system and benefit from the highly asynchronous inference system design and low bottleneck. The last line we set $max\_iter=20$.}
\label{tab:matrix}
\end{table}

\section*{Conclusion}

This work answers this missing piece of pipeline between DNNs built by machine learning practitioners and serves them efficiently with any hardware investment. We propose a solution to the complex allocation problem of multiple heterogenous DNNs in multiple heterogeneous GPUs. This flexibility relies on the formalism of the allocation matrix allowing data-parallelism, co-localization, and batch size optimization. 

To target efficiency, hundreds of allocation matrices are assessed before selecting the best one and instantiating it. And more, the inference system is built with many asynchronous processes to avoid overhead and accelerate the predictions. In our benchmarks, we observe the smart decision of our allocation optimizer. When the number of GPUs is superior to the number of DNNs, the heavier DNN are automatically multi-threaded to avoid bottleneck performance. On the opposite, when the number of GPUs is lower, we observe automatically co-localization and smaller batch size to fit all DNNs into the memory. 

Finally, each part of the proposed pipeline is well identified to guarantee easy code adaptation to facilitate introduction in current inference servers. For example, Object Detection and classification require different combination rules. To benefit from hardware-specific optimization, changing the inference framework requires localized updates into the \textit{predictor} process.

\subsubsection*{Acknowledgement}
We would like to thank TotalEnergies SE and its subsidiaries for allowing us to share this material and make available the needed resources.

\clearpage
\bibliography{bib_misc, bib_automl, bib_ensemble, bib_multi_goal, bib_compression, bib_baseline, bib_distrib}

\begin{thebibliography}{10}
\providecommand{\url}[1]{#1}
\csname url@samestyle\endcsname
\providecommand{\newblock}{\relax}
\providecommand{\bibinfo}[2]{#2}
\providecommand{\BIBentrySTDinterwordspacing}{\spaceskip=0pt\relax}
\providecommand{\BIBentryALTinterwordstretchfactor}{4}
\providecommand{\BIBentryALTinterwordspacing}{\spaceskip=\fontdimen2\font plus
\BIBentryALTinterwordstretchfactor\fontdimen3\font minus
  \fontdimen4\font\relax}
\providecommand{\BIBforeignlanguage}[2]{{%
\expandafter\ifx\csname l@#1\endcsname\relax
\typeout{** WARNING: IEEEtran.bst: No hyphenation pattern has been}%
\typeout{** loaded for the language `#1'. Using the pattern for}%
\typeout{** the default language instead.}%
\else
\language=\csname l@#1\endcsname
\fi
#2}}
\providecommand{\BIBdecl}{\relax}
\BIBdecl

\bibitem{useoverfit:1995}
P.~Sollich and A.~Krogh, ``Learning with ensembles: How overfitting can be
  useful.'' vol.~8, pp. 190--196, 01 1995.

\bibitem{uncertainty:2017}
B.~Lakshminarayanan, A.~Pritzel, and C.~Blundell, ``Simple and scalable
  predictive uncertainty estimation using deep ensembles,'' in \emph{NIPS},
  2017.

\bibitem{enscyb:2020}
S.~Haider, A.~Akhunzada, I.~Mustafa, T.~B. Patel, A.~Fernandez, K.-K.~R. Choo,
  and J.~Iqbal, ``A deep cnn ensemble framework for efficient ddos attack
  detection in software defined networks,'' \emph{IEEE Access}, vol.~8, pp.
  53\,972--53\,983, 2020.

\bibitem{enstim:2018}
S.~Pathak, X.~Cai, and S.~Rajasekaran, ``Ensemble deep timenet: An ensemble
  learning approach with deep neural networks for time series,'' in \emph{2018
  IEEE 8th International Conference on Computational Advances in Bio and
  Medical Sciences (ICCABS)}, 2018, pp. 1--1.

\bibitem{ensimg:2021}
M.~V. S.~d. Cea, D.~Gruen, and D.~Richmond, ``Pneumoperitoneum detection in
  chest x-ray by a deep learning ensemble with model explainability,'' in
  \emph{2021 IEEE 18th International Symposium on Biomedical Imaging (ISBI)},
  2021, pp. 1637--1641.

\bibitem{enssem:2021}
Q.~Sun and Z.~Ge, ``Deep learning for industrial kpi prediction: When ensemble
  learning meets semi-supervised data,'' \emph{IEEE Transactions on Industrial
  Informatics}, vol.~17, no.~1, pp. 260--269, 2021.

\bibitem{enstex:2020}
G.~Sun, J.~Liu, W.~Mengxue, W.~Zhongxin, Z.~Jia, and G.~Xiaowen, ``An ensemble
  classification algorithm for imbalanced text data streams,'' in \emph{2020
  IEEE International Conference on Artificial Intelligence and Computer
  Applications (ICAICA)}, 2020, pp. 1073--1076.

\bibitem{tritonserv}
T.~Xu, ``“deep into triton inference server: Bert practical deployment on
  nvidia gpu”,'' 2020, gPU Technology Conference.

\bibitem{ray}
P.~Moritz, ``{Ray: A Distributed Execution Engine for the Machine Learning
  Ecosystem },'' Electrical Engineering and Computer Sciences University of
  California at Berkeley, Tech. Rep., 08 2019.

\bibitem{tfserv}
C.~Olston, F.~Li, J.~Harmsen, J.~Soyke, K.~Gorovoy, L.~Lao, N.~Fiedel,
  S.~Ramesh, and V.~Rajashekhar, ``Tensorflow-serving: Flexible,
  high-performance ml serving,'' 2017, workshop on ML Systems at NIPS 2017.

\bibitem{torchserv}
Z.~DeVito, J.~Ansel, W.~Constable, M.~Suo, A.~Zhang, and K.~Hazelwood, ``Using
  python for model inference in deep learning,'' \emph{ArXiv}, vol.
  abs/2104.00254, 2021.

\bibitem{tensorrt}
P.~Davoodi, C.~Gwon, G.~Lai, and T.~Morris, ``“tensorrt inference with
  tensorflow”,'' 2019, gPU Technology Conference.

\bibitem{openvino}
Y.~Ge and M.~Jones, ``Inference with intel,'' 2018, aI DevCon 2018.

\bibitem{onnx:}
D.~D. Sarah~Bird, ``{ONNX},'' in \emph{Workshop NIPS2017}, 2017.

\bibitem{tflite:}
J.~Lee, N.~Chirkov, E.~Ignasheva, Y.~Pisarchyk, M.~Shieh, F.~Riccardi,
  R.~Sarokin, A.~Kulik, and M.~Grundmann, ``On-device neural net inference with
  mobile gpus,'' \emph{ArXiv}, vol. abs/1907.01989, 2019.

\bibitem{clipper:2017}
\BIBentryALTinterwordspacing
D.~Crankshaw, X.~Wang, G.~Zhou, M.~J. Franklin, J.~E. Gonzalez, and I.~Stoica,
  ``Clipper: A low-latency online prediction serving system,'' in \emph{14th
  {USENIX} Symposium on Networked Systems Design and Implementation ({NSDI}
  17)}.\hskip 1em plus 0.5em minus 0.4em\relax Boston, MA: {USENIX}
  Association, Mar. 2017, pp. 613--627. [Online]. Available:
  \url{https://www.usenix.org/conference/nsdi17/technical-sessions/presentation/crankshaw}
\BIBentrySTDinterwordspacing

\bibitem{dbms}
A.~Agrawal, R.~Chatterjee, C.~Curino, A.~Floratou, N.~Godwal, M.~Interlandi,
  A.~Jindal, K.~Karanasos, S.~Krishnan, B.~Kroth, J.~Leeka, K.~Park, H.~Patel,
  O.~Poppe, F.~Psallidas, R.~Ramakrishnan, A.~Roy, K.~Saur, R.~Sen, M.~Weimer,
  T.~Wright, and Y.~Zhu, ``Cloudy with high chance of {DBMS:} a 10-year
  prediction for enterprise-grade {ML},'' in \emph{10th Conference on
  Innovative Data Systems Research, {CIDR} 2020, Amsterdam, The Netherlands,
  January 12-15, 2020, Online Proceedings}, 2020.

\bibitem{WBF}
R.~Solovyev, W.~Wang, and T.~Gabruseva, ``Weighted boxes fusion: Ensembling
  boxes from different object detection models,'' \emph{Image Vis. Comput.},
  vol. 107, p. 104117, 2021.

\bibitem{worstfit}
{Michael J. Fischer.}, ``Thesis (ph. d.)--massachusetts institute of
  technology, dept. of mathematics,'' 1973.

\bibitem{pochelu:2021}
P.~Pochelu, S.~Petiton, and B.~Conche, ``A deep neural networks ensemble
  workflow from hyperparameter search to inference leveraging gpu clusters,''
  in \emph{Proceedings of the International Conference on High Performance
  Computing in Asia-Pacific Region, HPC Asia 2022, Virtual Event, January
  12-14, 2022}.\hskip 1em plus 0.5em minus 0.4em\relax {ACM}, 01 2022, p. to
  appear.

\bibitem{FastRCNN:2015}
\BIBentryALTinterwordspacing
R.~B. Girshick, ``Fast {R-CNN},'' \emph{CoRR}, vol. abs/1504.08083, 2015.
  [Online]. Available: \url{http://arxiv.org/abs/1504.08083}
\BIBentrySTDinterwordspacing

\end{thebibliography}

\bibliographystyle{IEEEtran}




\end{document}